\documentclass[a4paper,twocolumn,11pt,accepted=2025-09-30]{quantumarticle}
\pdfoutput=1
\usepackage[utf8]{inputenc}
\usepackage[english]{babel}
\usepackage{amsmath}
\usepackage{hyperref}
\usepackage{graphicx}
\usepackage{amssymb}
\usepackage{bm}

\begin{document}

\title{Monitored quantum transport:
full counting statistics of a quantum Hall interferometer}
\author{C. W. J. Beenakker}
\affiliation{Instituut-Lorentz, Universiteit Leiden, P.O. Box 9506, 2300 RA Leiden, The Netherlands}
\author{Jin-Fu Chen}
\affiliation{Instituut-Lorentz, Universiteit Leiden, P.O. Box 9506, 2300 RA Leiden, The Netherlands}
\affiliation{$\langle aQa^L\rangle$ Applied Quantum Algorithms Leiden, The Netherlands}
\date{April 2025}
\maketitle

\begin{abstract}
We generalize the Levitov-Lesovik formula for the probability distribution function of the electron charge transferred through a phase coherent conductor, to include projective measurements that monitor the chiral propagation in quantum Hall edge modes.  When applied to an electronic Mach-Zehnder interferometer, the monitoring reduces the visibility of the Aharonov-Bohm conductance oscillations while preserving the binomial form of the counting statistics, thereby removing a fundamental shortcoming of the dephasing-probe model of decoherence.
\end{abstract}

\section{Introduction}

The wave nature of a quantum particle manifests itself in the absence of which-path information: If the environment cannot detect which of two paths is taken by the particle, its wave nature allows for interference of the two probability amplitudes. Conversely, a which-path detector suppresses quantum interference, as has been demonstrated for electrons via Aharonov-Bohm conductance oscillations, mainly in the chiral transport regime of the quantum Hall effect \cite{Buk98,Spr00,Ji03,Lit07,Rou07,Rou08,Wei12,Hel15,Gur16,Jo21,Jo22}.

The microscopic theoretical modelling of these experiments is well developed \cite{See01,Cle04,Mar04,Suk07,Lev08,Neu08,You08,Sch11,Dre12,Idr18,Bel18}. Because of the fundamental nature of the loss of interference by which-path detection, there have also been attempts to arrive at a generic, model-independent description. The introduction of a dephasing probe is such an approach \cite{Chu05,Pil06,For07}, going back to early work by B\"{u}ttiker \cite{But88}. This approach opens up the system to an external reservoir that can absorb electrons and thereby provide which-path information. 

While the dephasing probe works well for the average current, for the current fluctuations it has a fundamental shortcoming noted by Marquardt and Bruder \cite{Mar04}: Occupation number fluctuations $\delta f$ in the external reservoir are introduced as a fluctuating $c$-number of time, without quantum statistical constraints \cite{note2}. As a consequence, the transferred charge no longer has the binomial distribution expected from Fermi statistics \cite{Bla00}. 

Here we present an alternative model-independent description of which-path detection that preserves the binomial nature of the charge transfer process. Drawing from concepts in quantum information processing \cite{NielsenChuang} we represent chiral transport by a quantum channel: a convex sum of Gaussian maps. Each term in the sum combines unitary evolution (described by a single-particle scattering matrix) with measurements that monitor the occupation number of certain modes.

Our central result is a generalization to monitored quantum transport of the celebrated Levitov-Lesovik formula \cite{Lev93,Lev96}, which expresses the distribution of transferred charge in terms of the scattering matrix of a phase coherent conductor. A fundamental consequence of the Levitov-Lesovik formula is that the charge transferred at zero temperature into a single outgoing mode by a voltage bias $V$ has a binomial distribution function,
\begin{equation}
P(Q)=\binom{N_{V}}{Q}T^Q(1-T)^{N_{V}-Q}.\label{binomial}
\end{equation}
The charge $Q$ transferred in the time $t_{\rm counting}$ is counted in units of the electron charge and $T\in(0,1)$ is the single-electron transfer probability. The number $N_{V}=(e/h)Vt_{\rm counting}$ is the number of incoming electrons during the time $t_{\rm counting}$, in a narrow energy interval $eV$ at the Fermi energy, sufficiently narrow that the energy dependence of $T$ can be neglected. The result \eqref{binomial} assumes $N_{V}\gg 1$, when the discreteness of the number of transferred charges no longer matters \cite{Sch07,Bed08,Avr08,Has08}.

As we will show in the following sections, the introduction of projective or weak measurements into the unitary dynamics has the effect of modifying the transfer probability, without changing the binomial nature of $P(Q)$. In the next section we first give the general representation of monitored chiral transmission as a convex-Gaussian quantum channel. The moment generating function of the transferred charge then follows, for projective measurements (Sec.\ \ref{sec_projective}) and for weak measurements (Sec.\ \ref{sec_weak}). The binomial statistics is derived in Sec.\ \ref{sec_binomial}. We apply the generalized Levitov-Lesovik formula to the quantum Hall interferometer in Sec.\ \ref{sec_QHI} and conclude in Sec.\ \ref{sec_conclude}. The appendix contains a generalization of Klich's trace-determinant relation \cite{Kli03} that we need for our analysis.

\section{Charge transfer statistics in a quantum channel}
\label{sec_projective}

The Levitov-Lesovik formula \cite{Lev93,Lev96} gives the moment generating function $F(\xi)=\langle e^{\xi Q}\rangle$ of the charge $Q$ of free electrons transferred through a conductor in the zero-frequency, long-time limit, under the assumption that the outgoing and incident density matrices are related by a unitary transformation, $\hat{\rho}_{\rm out}=\hat{U}\hat{\rho}_{\rm in}\hat{U}^\dagger$. We wish to generalize this to include projective measurements in addition to phase-coherent unitary evolution.

\subsection{Monitored chiral transmission as a convex-Gaussian quantum channel}

The most general relationship between $\hat{\rho}_{\rm out}$ and $\hat{\rho}_{\rm in}$ is the completely positive, trace preserving map of a quantum channel, represented by the operator sum \cite{NielsenChuang}
\begin{equation}
\hat{\rho}_{\rm out}=\sum_{\alpha} \hat{\cal K}_{\alpha}\hat{\rho}_{\rm in}\hat{\cal K}_{\alpha}^\dagger.\label{eq_operatorsum}
\end{equation}
The set of Kraus operators $\hat{\cal K}_\alpha$ sums to the identity operator,
\begin{equation}
\sum_\alpha \hat{\cal K}_\alpha^\dagger\hat{\cal K}_\alpha^{\vphantom{\dagger}}=\hat{I},\label{sumrule}
\end{equation}
to ensure that $\operatorname{Tr}\hat{\rho}_{\rm out}=\operatorname{Tr}\hat{\rho}_{\rm in}=1$.

\begin{figure}[tb]
\centerline{\includegraphics[width=0.8\linewidth]{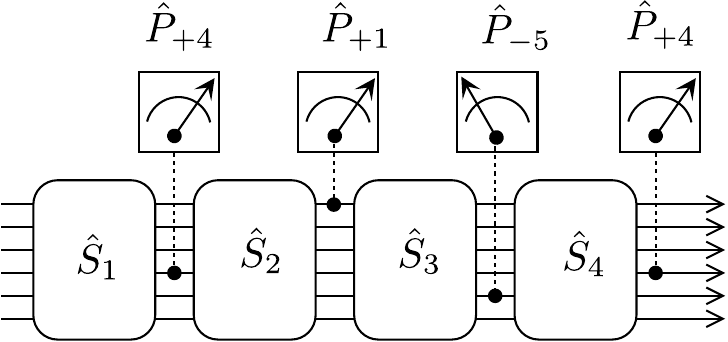}}
\caption{Schematic illustration of charge transfer via $N=6$ chiral modes, in which unitary propagation (scattering operators $\hat{S}_i$) alternates with $p=4$ projective measurements of the occupation number of specific modes. In this example the measurement outcomes are ``empty'' for the third measurement (of mode number $n=5$), and ``filled'' for the other three measurements.
}
\label{fig_channel}
\end{figure}

The monitored chiral transmission that we consider (see Fig.\ \ref{fig_channel}) consists of $p$ segments of unidirectional (=~chiral) propagation in $N$ modes, alternated by projective measurements of the occupation number of a specified mode. A measurement of the occupation number of mode $n$ that returns the value $0$ or $1$ has projector $a_n^{\vphantom{\dagger}}a_n^\dagger$ or $a_n^\dagger a_n^{\vphantom{\dagger}}$, respectively. The index $\alpha=(s_1,s_2,\ldots s_p)$ of the Kraus operator $\hat{\cal K}_\alpha$ labels the different measurement outcomes in the $p$ segments, with $s_i=+1$ if mode $n_i$ is filled and $s_i=-1$ if the mode is empty. The sum $\sum_\alpha$ thus runs over $2^p$ terms.

The measurements are alternated by unitary propagation, with scattering operators $\hat{S}_1,\hat{S}_2,\ldots \hat{S}_p$. Because the propagation is chiral, the scattering operators compose by multiplication, producing the Kraus operator
\begin{equation}
\begin{split}
&\hat{\cal K}_\alpha=\hat{P}_{s_p n_p}\hat{S}_p\cdots \hat{P}_{s_2 n_2}\hat{S}_2\hat{P}_{s_1 n_1}\hat{S}_1,\\
&\hat{P}_{+n}=a_n^\dagger a_n^{\vphantom{\dagger}},\;\;\hat{P}_{-n}=a_n^{\vphantom{\dagger}}a_n^\dagger.
\end{split}\label{Kalphadef}
\end{equation}

In second quantization the free-electron scattering operator $\hat{S}$ is the exponent of a quadratic form in the fermionic creation and annihilation operators,
\begin{equation}
\hat{S}=\exp\left(i\sum_{n,m=1}^{N}a_{n}^{\dagger}L_{nm}a_{m}\right)\equiv e^{ia^{\dagger}La}.
\end{equation}
We have collected the $N$ fermionic operators in vectors $a,a^\dagger$, contracted with an $N\times N$ Hermitian matrix $L=L^\dagger$.

The operator $\hat{S}$ corresponds in first quantization to a unitary scattering matrix $S=e^{iL}$ that relates incoming and outgoing single-particle states $|\psi_n\rangle=a_n^\dagger|0\rangle$,
\begin{align}
\hat{S}|\psi_n\rangle={}&e^{ia^\dagger L a}a_n^\dagger|0\rangle=\sum_{m=1}^N (e^{iL})_{mn}a^\dagger_m e^{ia^\dagger L a}|0\rangle\nonumber\\
={}&\sum_{m=1}^N (e^{iL})_{mn}a^\dagger_m |0\rangle=\sum_{m=1}^N S_{mn}|\psi_m\rangle.
\end{align}
In the second equality we used the identity
\begin{equation}
e^{a^\dagger Aa}a^\dagger_n e^{-a^\dagger Aa}=\sum_{m}(e^A)_{mn}a^\dagger_m.
\end{equation}

Substitution of Eq.\ \eqref{Kalphadef} into the operator sum \eqref{eq_operatorsum} defines the quantum channel as a convex sum of Gaussian maps \cite{Bra05,Mel13}. The sum rule \eqref{sumrule} follows from $\hat{S}_i^\dagger \hat{S}_i^{\vphantom{\dagger}}=\hat{I}$ and $\hat{P}_{+ n}^2+\hat{P}_{-n}^2=\hat{P}_{+ n}+\hat{P}_{-n}=\hat{I}$.

\subsection{Determinantal expression of the moment generating function}

The transferred charge $Q$ is measured in a subset of the outgoing modes, selected by the matrix
\begin{equation}
\bigl({\cal N}_{\rm out}\bigr)_{nm}=\begin{cases}
1&\text{if}\;\;n=m\in\{n^{\rm out}_{1},n^{\rm out}_2,\ldots n^{\rm out}_{N_{\rm out}}\},\\
0&\text{otherwise}.
\end{cases}
\end{equation}
The moment generating function is given by
\begin{subequations}
\label{FxiFxialpha}
\begin{align}
&F(\xi)\equiv\langle e^{\xi Q}\rangle=\operatorname{Tr}\hat{\rho}_{\rm out}e^{\xi a^\dagger {\cal N}_{\rm out}a}=\sum_\alpha F_{\bm\alpha}(\xi),\\
&F_{\alpha}(\xi)=\operatorname{Tr} \hat{\cal K}_{\alpha}\hat{\rho}_{\rm in}\hat{\cal K}_{\alpha}^\dagger e^{\xi a^\dagger {\cal N}_{\rm out} a},\\
&\hat\rho_{\rm in}=Z^{-1}e^{-\beta a^\dagger Ha},\;\;Z=\operatorname{Tr}e^{-\beta a^\dagger Ha}.
\end{align}
\end{subequations}

The incoming modes are in thermal equilibrium at inverse temperature $\beta=1/k_{\rm B}T_{\rm eq}$, with single-particle Hamiltonian $H$ that may have a different chemical potential for different modes, so that only some incoming modes may be occupied in an energy interval near the Fermi level.

Without any measurements there is only a single unitary Kraus operator
\begin{equation}
\hat{\cal K}_0=\hat{S}_p\cdots \hat{S}_2\hat{S}_1=e^{ia^\dagger L_pa}\cdots e^{ia^\dagger L_2a}e^{ia^\dagger L_1a}.
\end{equation}
The Levitov-Lesovik formula \cite{Lev93,Lev96} for the moment generating function $F_0$ then follows from Klich's trace-determinant relation \cite{Kli03},
\begin{align}
&\operatorname{Tr}e^{a^\dagger A_1a}e^{a^\dagger A_2a}\cdots e^{a^\dagger A_pa}=\operatorname{Det}\bigl(1+e^{A_1} e^{A_2}\cdots e^{A_p}\bigr),\label{Klichformula}\\
&\Rightarrow F_0(\xi)=\operatorname{Det}(1+e^{-\beta H})^{-1}\operatorname{Det}\bigl(1+{\cal S}_0 e^{-\beta H}{\cal S}_0^\dagger e^{\xi{\cal N}_{\rm out}}\bigr)\nonumber\\
&\qquad\qquad=\operatorname{Det}\bigl(1+{\cal N}_{\rm in}[{\cal S}_0^\dagger e^{\xi {\cal N}_{\rm out}}{\cal S}_0-1]\bigr).\label{LLformula}
\end{align}
We have defined
\begin{equation}
{\cal S}_0=S_p\cdots S_2S_1,\;\;{\cal N}_{\rm in}=(1+e^{\beta H})^{-1}.\label{S0def}
\end{equation}

To include the measurements we need to evaluate traces of operator products where Gaussian operators alternate with projectors onto filled or empty states. The required generalization of Klich's formula is derived in App.\ \ref{tracedetformula}. We first apply the anticommutator
\begin{equation}
\{ a_n^\dagger,a_n^{\vphantom{\dagger}}\}=1\Rightarrow \hat{P}_{+n}=1-\hat{P}_{-n}\label{anticommutator}
\end{equation}
to rewrite each trace as a sum of traces containing only projectors onto empty states. We then have the trace-determinant relation
\begin{align}
&\operatorname{Tr} a_1^{\vphantom{\dagger}} a_1^\dagger e^{a^\dagger A_1a}\cdots a_2^{\vphantom{\dagger}} a_2^\dagger e^{a^\dagger A_2 a}\cdots a_p^{\vphantom{\dagger}} a_p^\dagger e^{a^\dagger A_p a}=\nonumber\\
&\qquad=\operatorname{Det}\bigl(1+P_1 e^{A_1}P_2 e^{A_2}\cdots P_p e^{A_p}\bigr),\label{emptymodedet}
\end{align}
where $P_i$ is the $N\times N$ unit matrix with the $i,i$ element replaced by zero,
\begin{equation}
(P_i)_{nm}=\begin{cases}
1&\text{if} \;\;n=m\neq i,\\
0&\text{otherwise}.
\end{cases}\label{Qdef}
\end{equation}

The resulting expression for the moment generating function takes the form of a sum over $4^p$ determinants \cite{note1}, labeled by two strings of variables $\sigma=(\sigma_1,\sigma_2\ldots \sigma_p)$, $\tau=(\tau_1,\tau_2,\ldots\tau_p)$, with $\sigma_i,\tau_i\in\{0,1\}$:
\begin{subequations}
\label{Fxifull}
\begin{align}
F(\xi)={}&\sum_{\sigma,\tau}\left(\prod_{i=1}^p(-1)^{\sigma_i+\tau_i}(1+\sigma_i\tau_i)\right)\nonumber\\
&\times\operatorname{Det}\bigl(1+{\cal N}_{\rm in}[{\cal S}_{\tau}^\dagger e^{\xi {\cal N}_{\rm out}}{\cal S}_\sigma-1]\bigr),\\
\begin{split}
{\cal S}_{\sigma}={}&P^{\sigma_p}_{n_p}S_p\cdots P^{\sigma_2}_{n_2}S_2 P^{\sigma_1}_{n_1}S_1,\\
{\cal S}_{\tau}={}&P^{\tau_p}_{n_p}S_p\cdots P^{\tau_2}_{n_2}S_2 P^{\tau_1}_{n_1}S_1.
\end{split}
\end{align}
\end{subequations}
The matrices ${\cal S}_{\sigma},{\cal S}_\tau$ are $N\times N$ subunitary matrices, obtained as the product of unitary scattering matrices with certain rows and columns set to zero. The unitary matrix product ${\cal S}_0$ from Eq.\ \eqref{S0def} arises when $\sigma=(0,0,\ldots 0)$ (with the convention that $P_n^0$ is the unit matrix).

\section{Weak measurements}
\label{sec_weak}

So far we considered projective measurements onto a filled or empty mode. Let us generalize to a weak measurement, interpolating between the identity and a projection:
\begin{equation}
\begin{split}
&\hat{P}_{+n}(\varepsilon)=\delta\hat{I} +\varepsilon a_n^\dagger a_n^{\vphantom{\dagger}},\;\;\hat{P}_{-n}(\varepsilon)=\delta\hat{I}+\varepsilon a_n^{\vphantom{\dagger}}a_n^\dagger,\\
&\delta=\tfrac{1}{2}(\sqrt{2-\varepsilon^2}-\varepsilon),\;\;0\leq \varepsilon\leq 1.
\end{split}
\end{equation}
The Kraus operators are still trace preserving, because
\begin{equation}
\hat{P}_{+n}(\varepsilon)^2+\hat{P}_{-n}(\varepsilon)^2=\hat{I}.
\end{equation}

For $\varepsilon\neq 1$ the projector $\hat{P}_{\pm n}(\varepsilon)$ can be written as a Gaussian operator,
\begin{equation}
\begin{split}
&\hat{P}_{+n}(\varepsilon)=\delta e^{\gamma a_n^\dagger a_n^{\vphantom{\dagger}}},\;\;
\hat{P}_{-n}(\varepsilon)=\delta e^\gamma e^{-\gamma a_n^\dagger a_n^{\vphantom{\dagger}}},\\
&\gamma=\ln(1+\varepsilon/\delta),\;\;0\leq\varepsilon<1.
\end{split}
\end{equation}
The case $\varepsilon=1$ of a fully projective measurement can be reached at the end of the calculation by taking the limit.

The moment generating function now takes the form
\begin{equation}
\begin{split}
&F(\xi)=\sum_\alpha R^2_\alpha\operatorname{Det}\bigl(1+ {\cal N}_{\rm in}\bigl[\Xi_\alpha^\dagger e^{\xi{\cal N}_{\rm out}}\Xi_\alpha-1\bigr]\bigr),\\
&\alpha=(s_1,s_2,\ldots s_p),\;\;s_i\in\{+1,-1\},
\end{split}\label{Fepsweak}
\end{equation}
where we have defined
\begin{subequations}
\label{RalphaXialpha}
\begin{align}
&R_\alpha=c_{s_1}c_{s_2}\cdots c_{s_p},\;\;c_s=\begin{cases}
\delta&\text{if}\;\;s=+1,\\
\delta e^{\gamma}&\text{if}\;\;s=-1,
\end{cases}\\
&\Xi_\alpha= e^{s_p\gamma P_{n_p}}S_p\cdots e^{s_2\gamma P_{n_2}}S_2e^{s_1\gamma P_{n_1}} S_1.
\end{align}
\end{subequations}

For later use we note that the identity
\begin{equation}
e^{\xi {\cal N}_{\rm out}}=1+(e^{\xi}-1){\cal N}_{\rm out}
\end{equation}
allows us to rewrite Eq.\ \eqref{Fepsweak} as
\begin{subequations}
\label{Fxifull2}
\begin{align}
F(\xi)={}&\sum_{\alpha} R^2_\alpha  (\operatorname{Det}\Omega_{\alpha})\nonumber\\
&\times\operatorname{Det}\bigl(1+(e^\xi-1)\Omega_{\alpha}^{-1}{\cal N}_{\rm in}{\Xi}_{\alpha}^\dagger{\cal N}_{\rm out}{\Xi}_\alpha),\\
\Omega_{\alpha}={}&1+{\cal N}_{\rm in}({\Xi}_{\alpha}^\dagger {\Xi}_\alpha-1).
\end{align}
\end{subequations}
The sum rule \eqref{sumrule} implies that
\begin{equation}
F(0)=1\Rightarrow\sum_{\alpha}R^2_\alpha\operatorname{Det}\Omega_\alpha=1.\label{sumruleimplies}
\end{equation}

\section{Binomial distribution of single-mode charge transfer}
\label{sec_binomial}

We now restrict ourselves to the case $N_{\rm out}=1$ that only the charge transferred in a \textit{single} outgoing mode $|n_{\rm out}\rangle$ is counted. The incoming electrons at the Fermi level may occupy an arbitrary number of modes $N_{\rm in}$. At zero temperature they have the density matrix
\begin{equation}
\hat{\rho}_{\rm in}\bigl{|}_{E_{\rm F}}=|\psi_{\rm in}\rangle\langle\psi_{\rm in}|,\;\;|\psi_{\rm in}\rangle=a^\dagger_{n_{1}}a^\dagger_{n_2}\cdots a^\dagger_{n_{N_{\rm in}}}|0\rangle.
\end{equation}

Upon application of a bias voltage $V$, charge is transferred in an energy interval $eV$ around the Fermi energy $E_{\rm F}$. We assume that this interval is small enough that the energy dependence of the scattering matrices can be neglected. Since elastic scattering is energy dependent on the scale $\hbar/t_{\rm dwell}$, with $t_{\rm dwell}$ the typical dwell time of an electron in the scattering region, we need $eVt_{\rm dwell}/\hbar\ll 1$. The counting time $t_{\rm counting}$ is necessarily much larger than $t_{\rm dwell}$. In the long-time limit $eVt_{\rm counting}/\hbar\gg 1$ the cumulant generating function $C(\xi)$ can be evaluated at the Fermi level and the voltage bias enters as a prefactor \cite{Has08},
\begin{equation}
C(\xi)=N_V\ln F(\xi)\bigl{|}_{E_{\rm F}},\;\;N_V=(e/h)Vt_{\rm counting}.
\end{equation}

Because
\begin{equation}
e^{\xi a_{n_{\rm out}}^\dagger a_{n_{\rm out}}}=1+(e^\xi-1)a_{n_{\rm out}}^\dagger a_{n_{\rm out}},
\end{equation}
the moment generating function \eqref{FxiFxialpha} reduces in this single-mode case to
\begin{align}
&F(\xi)\bigl{|}_{E_{\rm F}}=\sum_\alpha \operatorname{Tr}\hat{\cal K}_{\alpha} \hat{\rho}_{\rm in}\bigl{|}_{E_{\rm F}}\hat{\cal K}_{\alpha}^\dagger [1+(e^\xi-1)a_{n_{\rm out}}^\dagger a_{n_{\rm out}}]\nonumber\\
&\Rightarrow \begin{cases}C(\xi)=N_V\ln\bigl(1+(e^\xi-1)T_{\rm eff}\bigr),\\
T_{\rm eff}=\sum_\alpha\langle\psi_{\rm in}|\hat{\cal K}_{\alpha}^\dagger a_{n_{\rm out}}^\dagger a_{n_{\rm out}} \hat{\cal K}_{\alpha}|\psi_{\rm in}\rangle.
\end{cases}\label{Cxibinomial}
\end{align}
Here we have applied the sum rule \eqref{sumrule}, restricted to the Fermi level, under the assumption that the measurements are quasi-elastic --- they do not significantly couple different energies.

The probability distribution function corresponding to Eq.\ \eqref{Cxibinomial} is binomial,
\begin{equation}
P(Q)=\binom{N_{V}}{Q}T_{\rm eff}^Q(1-T_{\rm eff})^{N_V-Q}.\label{PQeff}
\end{equation}
The measurements affect the transfer probability $T_{\rm eff}$ but not the binomial form of the charge transfer statistics. 

All of this applies to energy conserving scattering and measurement processes in the regime $k_{\rm B}T_{\rm eq}\ll eV$, $t_{\rm dwell}\ll \hbar/eV\ll t_{\rm counting}$. We did not specify the nature of the measurements, which may be more general than the single-mode projective measurement of Sec.\ \ref{sec_projective}. If we do specify to that case, we have an explicit expression for the effective transfer probability,
\begin{equation}
\begin{split}
&T_{\rm eff}=\sum_{\alpha}R_\alpha^2(\operatorname{Det}\Omega_{\alpha})\langle  n_{\rm out}|{\Xi}_\alpha\Omega_{\alpha}^{-1}{\cal N}_{\rm in}{\Xi}_{\alpha}^\dagger |n_{\rm out}\rangle,\\
&\bigl({\cal N}_{\rm in}\bigr)_{nm}=\begin{cases}
1&\text{if}\;\;n=m\in\{n_{1},n_2,\ldots n_{N_{\rm in}}\},\\
0&\text{otherwise},
\end{cases}
\end{split}
\label{Fxisinglemode}
\end{equation}
with $R_\alpha,\Omega_\alpha,\Xi_\alpha$ given by Eqs.\ \eqref{RalphaXialpha} and \eqref{Fxifull2}.

\section{Application to a quantum Hall interferometer}
\label{sec_QHI}

\begin{figure}[tb]
\centerline{\includegraphics[width=0.8\linewidth]{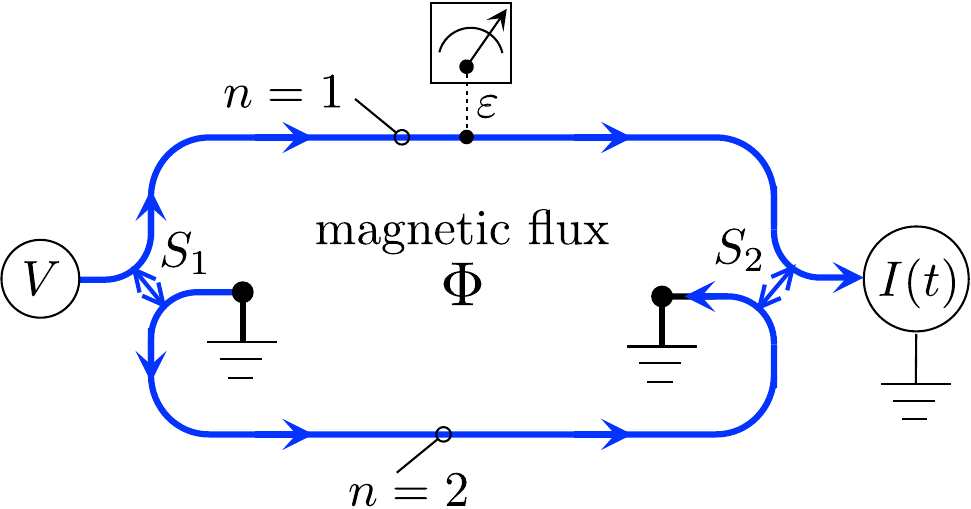}}
\caption{Schematic illustration of the Mach-Zehnder interferometer in a quantum Hall insulator. Two chiral edge modes ($n=1,2$) are coupled at a pair of beam splitters (scattering matrices $S_1$ and $S_2$). Charge is injected into the edge modes at the Fermi level by a voltage source $V$. A projective measurement (with probability $\varepsilon$) of the occupation of the $n=1$ edge mode provides ``which-path'' information that reduces the visibility in the outgoing current $I(t)$ of the Aharonov-Bohm oscillations as a function of the enclosed magnetic flux $\Phi$. For the time-averaged current the dephasing-probe model gives the same result as the projective measurement, for the current fluctuations only the projective measurement gives results consistent with binomial statistics.
}
\label{fig_MZ}
\end{figure}

We apply the general formulas to the quantum Hall interferometer of Fig.\ \ref{fig_MZ}, which is the electronic analogue of the optical Mach-Zehnder interferometer \cite{Spr00,Ji03,Lit07,Rou07,Rou08,Wei12,Hel15,Gur16,Jo21,Jo22}. Two chiral modes enclose a magnetic flux $\Phi$. A beam splitter distributes incoming electrons in a single mode over the two arms of the interferometer. The arms recombine at a second beam splitter, and a single outgoing mode is detected.

We parameterize the scattering matrices $S_1,S_2\in {\rm SU}(2)$ of the two beam splitters by 
\begin{equation}
S(\alpha,\alpha',\theta)=e^{i\alpha\sigma_z}e^{i\theta\sigma_y}e^{i\alpha'\sigma_z}.
\end{equation}
Phase shifts accumulated along the arms of the interferometer are absorbed in the parameters $\alpha,\alpha'$. These vary periodically with the enclosed flux, according to 
\begin{equation}
2(\alpha_1+\alpha'_2)=\phi_0+e\Phi/h,
\end{equation}
with $\phi_0$ a magnetic field independent offset.

Since the transferred charge is counted in a single mode, we can apply Eq.\ \eqref{Fxisinglemode}. The resulting probability distribution function has the binomial form \eqref{PQeff} with effective transfer probability
\begin{align}
T_{\rm eff}={}&\tfrac{1}{2}  (1-\cos 2 \theta_1 \cos 2 \theta_2)\nonumber\\
&+\tfrac{1}{2}(1-\varepsilon^2) \cos(\phi+e\Phi/h)\sin 2\theta_1\sin 2\theta_2.
\end{align}
The enclosed-flux dependent oscillations vanish in the limit $\varepsilon\rightarrow 1$, as it should be, while the binomial statistics persists for any $\varepsilon$.

To compare with the dephasing-probe calculations \cite{Mar04,Chu05,Pil06,For07}, we take two equal 50/50 beam splitters ($\theta_1=\theta_2=\pi/4$), and maximal dephasing ($\varepsilon=1$), when the mean and variance of the transferred charge are given by
\begin{equation}
\left\langle Q\right\rangle =\frac{1}{2}N_V,\;\;\operatorname{Var}(Q)=\frac{1}{4}N_V.
\end{equation}
The Fano factor $\operatorname{Var}(Q)/\left\langle Q\right\rangle =1/2$, characteristic of an unbiased binomial process. The dephasing probe, instead, gives for this case \cite{Mar04,Chu05,Pil06,For07}
\begin{equation}
\left\langle Q\right\rangle =\frac{1}{2}N_V,\;\;\operatorname{Var}(Q)=\frac{1}{8}N_V,
\end{equation}
so Fano factor $1/4$, inconsistent with binomial statistics.

\section{Conclusion}
\label{sec_conclude}

The monitored quantum transport description of decoherence that we have developed is an abstraction of a complex microscopic problem \cite{See01,Cle04,Mar04,Suk07,Lev08,Neu08,You08,Sch11,Dre12,Idr18,Bel18}. In a typical experiment there may be several sources of dephasing, such as fluctuations in the electromagnetic environment, coupling to lattice vibrations, and electron-electron interactions. A generic aspect is that these are mechanisms for which-path information, and it is that abstraction that is captured by the projective measurements.

As emphasised in one of the first experiments on the quantum Hall interferometer \cite{Ji03}, the visibility of quantum interference effects can also be reduced by phase averaging, due to a finite temperature or due to variations in the system on the time scale of the experiment. Such phase averaging is altogether different from dephasing \cite{Mar04}, in particular, phase averaging causes deviations from binomial statistics --- in contrast to dephasing. Monitored quantum transport allows for a study of dephasing without the confounding effects of phase averaging.

Our focus here has been on the derivation of the generalized Levitov-Lesovik formula (Eqs.\ \eqref{Fxifull} and \eqref{Fxifull2} for projective and weak measurements, respectively), and the demonstration of binomial statistics. Follow-up work could be motivated by noting that the original Levitov-Lesovik formula \cite{Lev93,Lev96} has opened up the study of charge transfer statistics to scattering-matrix based methods, such as random-matrix theory \cite{Bla00}. 

In the broader context of monitored quantum circuits, random-matrix models have recently been used to study the limit of a large number of weak measurements \cite{Bul24,Ger24,Xia25,Bee25}, and one could consider a similar application to a study of the effect of dephasing on shot noise in a disordered chiral system, such as the \textit{p-n} junction in graphene \cite{Jo21,Jo22}: In a junction of length $L$ one could introduce a series of weak measurements of strength $\varepsilon^2\propto\delta L$ spaced by $\delta L$, and take the limit $\varepsilon\rightarrow 0$, $L/\delta L\rightarrow\infty$ at fixed $\varepsilon^2 L/\delta L$ to model a continuous spatially distributed dephasing process.

Another direction of future research is to include feedback based on the measurement outcome, so that we find the charge transfer statistics conditioned on a set of projective measurements. The formalism developed here is general enough to allow for that.

Finally, the original Levitov-Lesovik formula applies also if the transport is not chiral, in particular, it can be applied to the study of current fluctuations in the presence of Anderson localization by static disorder \cite{Bla00}. Monitored quantum transport of disordered conductors has been studied recently \cite{Jia22,Pan24,Gur25} and it would be of interest to derive the generalized Levitov-Lesovik formula without the chirality assumption.

\acknowledgments

Results from App.\ \ref{tracedetformula} were developed in discussions at the \href{https://quantumcomputing.stackexchange.com/q/41449/2555}{quantumcomputing} and \href{https://mathoverflow.net/q/488922/11260}{mathoverflow} Q\&A sites. We thank Max Alekseyev and Fred Hucht for their input.

This project was supported by the Netherlands Organisation for Scientific Research (NWO/OCW), as part of Quantum Limits (project number SUMMIT.1.1016). J.F.C. also acknowledges the support received from the European Union's Horizon Europe research and innovation programme through the ERC StG FINE-TEA-SQUAD (Grant No. 101040729).

The views and opinions expressed here are solely those of the authors and do not necessarily reflect those of the funding institutions. None of the funding institutions can be held responsible for them.
\appendix

\section{Determinantal expression of the trace of projections alternating with Gaussian operators}
\label{tracedetformula}

We seek to generalize Klich's trace-determinant relation \eqref{Klichformula} to include projectors in the product of Gaussian operators. We first consider the case that all projectors are onto empty states, leading to Eq.\ \eqref{emptymodedet} in the main text.

\subsection{Projection onto empty states}

The projector $a_n^{\vphantom{\dagger}}a^\dagger_n$ onto the empty mode $n$ can be written as the limit of a Gaussian operator \cite{Kni01},
\begin{equation}
a_n^{\vphantom{\dagger}}a^\dagger_n=\lim_{f\rightarrow\infty}e^{-f a_n^\dagger a_n^{\vphantom{\dagger}}},\label{emptystatelimit}
\end{equation}
so that the trace has the expression
\begin{widetext}
\begin{align}
F_{\rm empty}={}&\lim_{f_{1},f_{2},\ldots f_{d}\rightarrow\infty}\operatorname{Tr}\bigl(e^{- f_{1} a_{1}^\dagger a^{\vphantom{\dagger}}_{1}}\hat{M}_1e^{- f_{2} a_{2}^\dagger a^{\vphantom{\dagger}}_{2}}\hat{M}_2\cdots e^{- f_{d} a_{d}^\dagger a^{\vphantom{\dagger}}_{d}}\hat{M}_d\bigr)\nonumber\\
={}&\lim_{f_{1},f_{2},\ldots f_{d}\rightarrow\infty}\operatorname{Tr}\bigl(e^{- a^\dagger Q_{1}(f_1)a}\hat{M}_1e^{- a^\dagger Q_{2}(f_2)a}\hat{M}_2\cdots e^{- a^\dagger Q_{d}(f_d)a}\hat{M}_d\bigr).
\end{align}
We have defined
\begin{equation}
[Q_{ n}(z)]_{ij}=\begin{cases}
z &\text{if}\;\;i=j=n,\\
0&\text{otherwise}.
\end{cases}
\end{equation}

The limit can be taken by first rewriting the trace over $N$ modes in Fock space as a determinant of an $N\times N$ matrix \cite{Kli03},
\begin{align}
F_{\rm empty}={}&\lim_{f_{1},f_{2},\ldots f_{d}\rightarrow\infty}\operatorname{Det}\bigl(1+e^{-Q_{1}(f_1)}e^{A_1}e^{-Q_{2}(f_2)}e^{A_2}\cdots e^{-Q_{d}(f_d)}e^{A_d}\bigr)\nonumber\\
={}&\operatorname{Det}\bigl(1+P_{1}e^{A_1}P_{2}e^{A_2}\cdots P_{d}e^{A_d}\bigr),\label{Femptyresult}
\end{align}
with $P_n=\lim_{f\rightarrow\infty}e^{-Q_n(f)}$ the matrix defined in Eq.\ \eqref{Qdef}. We have thus arrived at the result \eqref{emptymodedet}.
\end{widetext}

\subsection{Projection onto filled states}

If all projections are onto filled states one has the trace
\begin{equation}
F_{\rm filled}=\operatorname{Tr}(a^\dagger_{1}a^{\vphantom{\dagger}}_{1}  \hat{M}_1 a^\dagger_{2}a^{\vphantom{\dagger}}_{2}  \hat{M}_2\cdots  a^\dagger_{d}a^{\vphantom{\dagger}}_{d}{\hat{M}}_d).
\end{equation}
The filled-state projector can also be written as the limit of a Gaussian operator \cite{Kni01},
\begin{equation}
a_n^\dagger a_n^{\vphantom{\dagger}}=\lim_{f\rightarrow\infty}e^{-f}e^{f a_n^\dagger a^{\vphantom{\dagger}}_n}.\label{filledstatelimit}
\end{equation}

Application of Klich's formula \eqref{Klichformula} gives
\begin{widetext}
\begin{equation}
F_{\rm filled}=\lim_{f_{1},f_{2},\ldots f_{d}\rightarrow\infty}e^{-\sum_n f_n}
\operatorname{Det}\bigl(1+e^{Q_{1}(f_1)}e^{A_1}e^{Q_{2}(f_2)}e^{A_2}\cdots e^{Q_{d}(f_d)}e^{A_d}\bigr).
\end{equation}
To carry out the limit we factorize the determinant,
\begin{align}
F_{\rm filled}={}&\lim_{f_{1},f_{2},\ldots f_{d}\rightarrow\infty}e^{-\sum_n f_n}\operatorname{Det}\bigl(e^{Q_{1}(f_1)}e^{A_1}e^{Q_{2}(f_2)}e^{A_2}\cdots e^{Q_{d}(f_d)}e^{A_d}\bigr)\nonumber\\
&\times\operatorname{Det}\bigl(1 + (e^{Q_{1}(f_1)}e^{A_1}e^{Q_{2}(f_2)}e^{A_2}\cdots e^{Q_{d}(f_d)}e^{A_d})^{-1}\bigr)\nonumber\\
={}&\operatorname{Det}\left(\textstyle{\prod_n} e^{A_n}\right)\lim_{f_{1},f_{2},\ldots f_{d}\rightarrow\infty}\operatorname{Det}\bigl(1+ e^{-Q_{1}(f_1)}e^{-A_1^\top}e^{-Q_{2}(f_2)}e^{-A_2^\top}\cdots e^{-Q_{d}(f_d)}e^{-A_d^\top}\bigr)\nonumber\\
={}&e^{\sum_n\operatorname{Tr}A_n}\operatorname{Det}\bigl(1+ P_1e^{-A_1^\top} P_2e^{-A_2^\top}\cdots P_de^{-A_d^\top}\bigr).\label{Ffilledresult}
\end{align}
Note that the $A$-matrices are transposed, not conjugate-transposed.
\end{widetext}

\subsection{Mixed empty/filled projections}
\label{tracedetformulac}

In the main text we substitute $a_n^\dagger a_n^{\vphantom{\dagger}}=1-a_n^{\vphantom{\dagger}}a_n^\dagger$ to express a trace with $p$ filled-state projections as a sum of $2^p$ determinants of the form \eqref{Femptyresult}. This exponential scaling can be avoided, at the expense of a more complicated formula, by the following steps.

Consider for example the mixed trace
\begin{equation}
F_{\rm mixed}=\operatorname{Tr}(a^\dagger_{1}a^{\vphantom{\dagger}}_{1}  \hat{M}_1 a^{\vphantom{\dagger}}_{2}a^\dagger_{2}  \hat{M}_2  a^\dagger_{1}a^{\vphantom{\dagger}}_{1}{\hat{M}}_3).
\end{equation}
We permute rows or columns so that the filled-state projections are all on the same mode (number 1 in this case). The empty-state projections can be on arbitrary modes.

We replace the projectors by limits of Gaussian operators, via Eqs.\ \eqref{emptystatelimit} and \eqref{filledstatelimit}, and apply the trace-determinant relation \eqref{Klichformula},
\begin{widetext}
\begin{align}
F_{\rm mixed}={}&\lim_{f_{1},f_{2}, f_{3}\rightarrow\infty}e^{-f_1-f_3}\operatorname{Tr}\bigl(e^{a^\dagger Q_{1}(f_1)a}\hat{M}_1e^{- a^\dagger Q_{2}(f_2)a}\hat{M}_2 e^{a^\dagger Q_{2}(f_3)a}\hat{M}_3\bigr)\nonumber\\
={}&\lim_{f_{1},f_{2}, f_{3}\rightarrow\infty}e^{-f_1-f_3}\operatorname{Det}\bigl(1+e^{Q_{1}(f_1)}X(f_2) e^{Q_{1}(f_3)}e^{A_3}\bigr),\;\;X(f_2)=e^{A_1}e^{-Q_{2}(f_2)}e^{A_2}.
\end{align}
\end{widetext}
The limit is taken of a multinomial in the variables $z_i=e^{-f_i}$, where each $z_i$ only appears with a power of zero or one. We can thus take the limit $f_i\rightarrow\infty\Rightarrow z_i\rightarrow 0$ in any order.

By taking first the filled-state limits $f_1,f_3\rightarrow\infty$, keeping the empty-state variable $f_2$ finite, we avoid a non-invertible $X$. We can then take the same steps as in the previous sub-section,
\begin{align}
F_{\rm mixed}={}&\lim_{f_2\rightarrow\infty}\bigl(\operatorname{Det}X(f_2)e^{A_3}\bigr)\nonumber\\
&\times\operatorname{Det}\bigl(1+P_1X^{\top}(f_2)^{-1} P_1 e^{-A_3^\top}\bigr).
\end{align}

At this stage we apply the identity \cite{Hucht}
\begin{align}
&\operatorname{Det}(A_1A_2\cdots A_d)\operatorname{Det}\bigl(1+P_1 A_1^{-1} P_1 A_2^{-1} \cdots P_1 A_d^{-1}\big)\nonumber\\
&=\left(\prod_{i=1}^d [A_i]_{11}\right)\operatorname{Det}\bigl(1+S_{A_d}\cdots S_{A_2} S_{A_1}\bigr),\label{Srelation2}
\end{align}
which holds for any set of invertible square matrices $A_i$ with nonzero 1,1 elements  (see App.\ \ref{tracedetformulad} for a derivation). The matrix $S_A$ is the Schur complement of $A$ with respect to the 1,1 element. Notice that the order in which the matrices $A$ and $S$ appear is inverted. 

We thus have
\begin{equation}
F_{\rm mixed}=\lim_{f_2\rightarrow\infty}[X(f_2)]_{1,1}[e^{A_3}]_{1,1}\operatorname{Det}\bigl(1+S^\top_{e^{A_3}}S^\top_{X(f_2)}\bigr).
\end{equation}
Because the Schur complement of $X$ remains well-defined if $X$ becomes singular, provided the 1,1 element remains nonzero, we can now take the limit $f_2\rightarrow\infty$, to arrive at
\begin{equation}
F_{\rm mixed}=[e^{A_1}P_2e^{A_2}]_{1,1}[e^{A_3}]_{1,1}\operatorname{Det}\bigl(1+S_{e^{A_1}P_2e^{A_2}}S_{e^{A_3}}\bigr).
\end{equation}

In this way we have expressed the mixed projector onto two filled and one empty mode in terms of a single determinant. The alternative approach, using the anticommutator, would have produced a sum of four determinants. Because the Schur complement expressions are somewhat less transparent, we use the alternative approach in the main text.

\subsection{Derivation of the Schur complement identity \eqref{Srelation2}}
\label{tracedetformulad}

Consider an $N\times N$ matrix with a bordered structure,
\begin{equation}
A=\begin{pmatrix}
a&\bm{b}^\top\\
\bm{c}&D
\end{pmatrix},
\end{equation}
where $a=A_{1,1}$ is a scalar, $\bm{b}$ and $\bm{c}$ are $(N-1)$-dimensional column vectors, and $D$ is a $(N-1)\times (N-1)$ matrix. We assume $a\neq 0$. The Schur complement $S_A$ of $A$ with respect to its 1,1 element is defined by
\begin{equation}
S_A=D-a^{-1}\bm{c}\bm{b}^\top.
\end{equation}
The determinants are related by
\begin{equation}
\operatorname{Det}A=A_{11}\operatorname{Det}S_A.\label{DetASrelation}
\end{equation}

The corresponding structure of the inverse of $A$ is
\begin{equation}
A^{-1}=\begin{pmatrix}
a^{-1}+a^{-2}\bm{b}^\top S_A^{-1}\bm{c}&-a^{-1}\bm{b}^\top S_A^{-1}\\
-a^{-1}S_A^{-1}\bm{c}&S_A^{-1}
\end{pmatrix}.
\end{equation}
Multiplication from the left with the matrix $P_1=\operatorname{diag}(0,1,1\ldots 1)$ zeroes out the first row. We alternate matrices $A_1^{-1},A_2^{-1},\ldots A_d^{-1}$ with the same projector $P_1$,
\begin{widetext}
\begin{align}
P_1 A_1^{-1} P_1 A_2^{-1} \cdots P_1 A_d^{-1}={}&\begin{pmatrix}
0&0\\
-a_1^{-1}S_{A_1}\bm{c}_1&S_{A_1}^{-1}
\end{pmatrix}
\begin{pmatrix}
0&0\\
-a_2^{-1}S_{A_2}\bm{c}_2&S_{A_2}^{-1}
\end{pmatrix}
\cdots
\begin{pmatrix}
0&0\\
-a_d^{-1}S_{A_d}\bm{c}_d&S_{A_d}^{-1}
\end{pmatrix}\nonumber\\
={}&\begin{pmatrix}
0&0\\
\bm{x} & S_{A_1}^{-1} S_{A_2}^{-1}\cdots S_{A_d}^{-1}
\end{pmatrix},
\end{align}
for some unspecified vector $\bm{x}$.

We can now compute the determinant
\begin{align}
\operatorname{Det}\bigl(1+P_1 A_1^{-1} P_1 A_2^{-1} \cdots P_1 A_d^{-1}\big)={}&\operatorname{Det}\begin{pmatrix}
1&0\\
\bm{x}&1+S_{A_1}^{-1} S_{A_2}^{-1}\cdots S_{A_d}^{-1}
\end{pmatrix}\nonumber\\
={}&\operatorname{Det}\bigl(1+S_{A_1}^{-1} S_{A_2}^{-1}\cdots S_{A_d}^{-1}\bigr).\label{Srelation1}
\end{align}
\end{widetext}
Multiplication by $\operatorname{Det}(A_1A_2\cdots A_d)$ and use of Eq.\ \eqref{DetASrelation} gives the desired relation \eqref{Srelation2}.


\begin{thebibliography}{99}
\bibitem{note2} We do wish to emphasise that this shortcoming of the dephasing-probe
model only applies to the current fluctuations, not to the time-averaged
current and hence not to the conductance. Most applications of the dephasing-probe model only address the conductance, the reservoir fluctuations play no role.
\bibitem{note1} The sum \eqref{Fxifull} over $4^p$ determinants can be reduced to a sum over $2^p$ determinants by avoiding the substitution \eqref{anticommutator}, and directly evaluating combinations of projectors onto filled and empty states, see App.\ \ref{tracedetformulac}. The resulting expressions, in terms of the Schur complements of the scattering matrices, are more complicated to analyze, so we do not take that approach.
\bibitem{Hucht} Eq.\ \eqref{Srelation2} was pointed out to us by Fred Hucht.
\end{thebibliography}
\end{document}